# Coupled electromigration-nanoindentation study on dislocation nucleation in $SrTiO_3$

Chukwudalu Okafor[1]*, Ahmad Sayyadi-Shahraki[2], Sebastian Bruns[2], Till Frömling[3], Pierre Hirel[4], Phillipe Carrez[4], Karsten Durst[2], Xufei Fang[1]*

[1]Institute for Applied Materials, Karlsruhe Institute of Technology, Karlsruhe, Germany

[2]Department of Materials and Earth Sciences, Technical University of Darmstadt, Germany

[3]Institute for Energy Technology IET-1, Forschungszentrum Jülich, Institute for Energy Technology, IET-1, 52428, Jülich Germany

[4]Université Lille, CNRS, INRAE, Centrale Lille, UMR 8207 - UMET - Unité Matériaux et Transformations, France

*Corresponding authors: chukwudalu.okafor2@kit.edu (C.O.); xufei.fang@kit.edu (X.F.)

## Abstract

Modern functional oxides are mainly engineered by doping, essentially by tuning the defect chemistry. Recent studies suggest that dislocations offer a new perspective for enhancing the mechanical and physical properties of ceramic oxides. This raises the question regarding the interaction between dislocations and point defects in ceramics. Here, we report the impact of defect chemistry on the mechanical response of single-crystal strontium titanate, a prototype perovskite oxide. We demonstrate that electric field-induced stoichiometry polarization alters the defect chemistry, primarily by tuning oxygen vacancies, resulting in a distinct difference in the maximum shear stresses for dislocation nucleation, as experimentally observed and corroborated by molecular dynamic simulation. The impact of indenter tip size and geometry on the dislocation nucleation behavior in samples with different point defect concentrations in ceramics is further elucidated. Similar to the electromigration findings, acceptor-doped $SrTiO_3$ tends to promote dislocation nucleation due to the abundance of oxygen vacancies. These findings shed new light on the interaction between dislocations and point defects in oxides. They may pave the road for assessing the stability of the next-generation functional ceramics engineered by dislocations.

## 1. Introduction

Oxide perovskites are essential in electroceramic devices with a wide range of applications, including capacitors, actuators, and oxygen sensors [1, 2]. The movement of oxygen ions (particularly at elevated temperatures [3]) underscores their applicability as electrode material for e.g., solid oxide fuel cell application [4, 5]. Recent technological advancements in oxide electronics demand miniaturization. As devices become smaller, the effective volume of defects increases concerning the active device area. Certain components of these devices, for instance ferroelectric and dielectric, are commonly sandwiched between electrodes as capacitors or, in some cases, as thin films with lattice-mismatched interfaces, leading to the formation of dislocations. Although ceramics are generally termed as "brittle", dislocations are almost unavoidable in bulk ceramics processing as they are present even at elevated temperatures (for instance, $SrTiO_3$ up to $0.8T_m$ [6]). These dislocations are not entirely "unwanted" in functional ceramics for electroceramic applications, as recent studies have shown that dislocation engineering may hold promising potential to enhance the functional properties [7-10] and mechanical properties [11, 12] in functional ceramics. Particularly, dislocations were found to form "conducting nanowires" in some cases, which enhance the electrical conductivity of conventionally insulating $Al_2O_3$ and yttria-stabilized zirconia (YSZ) [13-16]. In many applications, miniaturized devices experience gradual or instantaneous temperature spikes, which could alter the defect chemistry (vacancy formation [2, 17-19]). Eventually, the point defects interact with the dislocations, which may further influence the dislocation behavior (dislocation nucleation, multiplication, and motion [20]). Hence, understanding the interactions between point defects and dislocations is critical for the next-generation functional oxides.

Various attempts have been made to understand the interaction between point defects and dislocations in ceramics, with an overview dating back almost 50 years [21]. With renewed interest in dislocation studies in ceramics, attempts to understand the impact of point defects on dislocation-based plasticity and cracking via point defect engineering, such as non-stoichiometry [22, 23] and thermal treatment [24], have resurfaced by probing the impact of point defects on dislocation mechanics through mechanical deformation. For instance, Nakamura et al. [23] examined the effect of stoichiometry change (i.e., by changing the Sr/Ti ratio in the starting powders for growing the single crystals) on the plastic deformation of single-crystal $SrTiO_3$. They observed a slightly higher degree of plasticity during uniaxial bulk compression for non-stoichiometric $SrTiO_3$, which was attributed to a lower overall vacancy concentration. Subsequently, Fang et al. [22] investigated the same samples used by Nakamura et al. [23] via nanoindentation. They demonstrated that non-stoichiometric $SrTiO_3$ crystals exhibit easier dislocation nucleation, as indicated by the lower nanoindentation pop-in stresses. However, the dislocation motion in such samples was more difficult. The proposed hypothesis was that a higher concentration of oxygen vacancies may have resulted in a solute drag-like effect[22, 25]. These works represent a preliminary experimental attempt to probe the dislocation mechanics of $SrTiO_3$ with a pre-engineered defect state. Additionally, the interaction between dislocations and point defects has been investigated through molecular dynamics (MD) simulations of nanoindentation in the presence of oxygen vacancies on single-crystal Fe [26]. Njeim et al. [26] observed that a relatively lower load is required to nucleate dislocations near oxygen vacancies during nanoindentation simulations.

However, it is still unclear, particularly in oxides, which point defect species (cation and/or anion vacancies) contribute to the observed dislocation behavior. There has been a lack of direct experimental evidence on the dislocation mechanisms for decoupling the impact of defect species, especially anions ($V_O^{\cdot\cdot}$) and cations ($V_{Sr}''$) vacancies present in $SrTiO_3$. In this work, we aim to address the following question: Can we isolate individual defect species, cation, and anion vacancies in oxides using $SrTiO_3$ as a model perovskite oxide? If possible, how will each defect species impact the dislocation nucleation in $SrTiO_3$? To address these questions, we adopt the approach of electric field-induced stoichiometry polarization, i.e., the build-up of defect concentration profile, which is commonly induced under an electric field with partially or entirely charge carrier-blocking electrodes [27].

For oxides such as $SrTiO_3$, the stoichiometry polarization is generally associated with the electromigration of oxygen vacancies and their accumulation at the cathode (hence depletion at the anode) ensured by ionic-blocking electrodes [28, 29]. Hence, ionic conduction is suppressed, while electron-hole transport dominates the observed conductivity during the electromigration of oxygen vacancies. Aside from the conductivity, stoichiometry polarization is also accompanied by electrocoloration (i.e., changes in optical appearances), which has been reported on single-crystal undoped [27, 28, 30] and Fe-doped $SrTiO_3$ [31]. The color fronts were observed to emerge from the electrodes (at the edges of the crystal), propagating into the bulk under an applied DC field [28, 31]. The optical change is direct evidence of the migration of oxygen vacancies, as the charge state of cation elements changes to accommodate for the polarized vacancy concentration.

In this work, we performed nanoindentation testing coupled with electromigration experiment on an undoped single-crystal $SrTiO_3$ sample to gain insight into the influence of point defects on dislocation nucleation. We relied on the influence of stoichiometry polarization-induced electromigration to generate an oxygen vacancy concentration profile within a single sample. Additionally, MD (molecular dynamics) simulation was employed to corroborate the experimental observations.

## 2. Experimental and simulation

### 2.1. Materials selection and surface characterization

The behavior of point defects in $SrTiO_3$ has been extensively studied, making $SrTiO_3$ a prime candidate for this investigation. Verneuil-grown, nominally undoped single-crystal $SrTiO_3$ sample (Alineason Materials Technology, GmbH, Frankfurt Am Main, Germany), with dimensions of 10 x 10 x 1 $mm^3$, was used for the electromigration and nanoindentation testing.

### 2.2. Electromigration for tuning the oxygen vacancy concentrations

We first performed a stoichiometry polarization-induced electromigration test on the undoped $SrTiO_3$ sample to generate a concentration profile of oxygen vacancy before nanoindentation testing. The electromigration test was performed using a micro-contact setup under an applied electric field as illustrated in **Fig. 1A** Platinum (Pt) electrodes were sputtered on two opposite ends of the same surface (dimensions 10 x 3 x 1 $mm^3$) using a shadow mask to create an electronic conductive layer (**Fig. 1B**). The electrodes were contacted with a DC electric field of 85 V (Keithley Instruments, Ohio, USA) via tungsten carbide (WC) tips. The sample was heated up to 350 ± 5 °C (623 ± 5 K) on a crucible (Linkam Scientific Instruments, Tadham UK). At this temperature, there is enhanced mobility of charge carriers (primarily oxygen vacancies) [32]. A commercially available thermocouple was used to ascertain the sample surface temperature. The DC electric field was retained for ~1 hour [27].

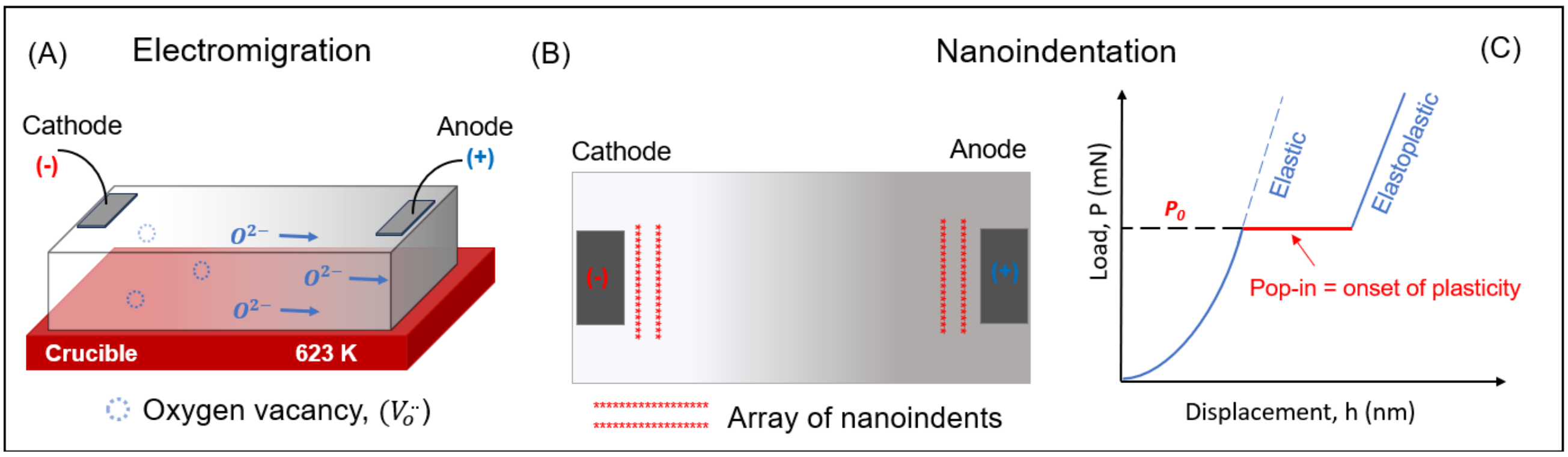

***Fig. 1:*** *(A) Schematic illustration of the electromigration experiment showing the migration of oxygen ions to the anode and retained oxygen vacancies on the cathode (blue circles). (B) Illustration of a* $SrTiO_3$ *crystal after an electromigration experiment. The red asterisk represents the position of nanoindents near the cathode and anode regions. Visible optical changes (electrocoloration) due to the migration of* $V_O^{\cdot\cdot}$ *are also depicted. (C) A representative nanoindentation load-displacement plot illustrates the pop-in event (on-set of plasticity) with a corresponding elastic Hertzian fit, as indicated by the blue dashed line.*

### 2.3. Nanoindentation experiment

After electromigration, nanoindentation was carried out to observe the impact of the $V_O^{\cdot\cdot}$ gradient on dislocation nucleation. Nanoindentation was performed via the continuous stiffness measurement (CSM) mode using a spherical indenter tip (with an effective tip radius of 1.3 μm) and a constant strain rate of 0.05 $s^{-1}$ on a G200 nanoindenter (KLA Instruments, California, USA) in the load-controlled mode. The spherical tip ensures a larger stressed volume (compared to the Berkovich indenter tip) with a higher probability of probing more point defects ($V_O^{\cdot\cdot}$). A maximum tip displacement of 200 nm into the sample was set for all nanoindentation tests, with a harmonic displacement and frequency target of 2 nm and 45 Hz, respectively. With the variation in the $V_O^{\cdot\cdot}$ concentration ensured by electromigration, arrays of nanoindents were placed just next to the Pt electrodes (cathode: $V_O^{\cdot\cdot}$ Enriched region; anode: $V_O^{\cdot\cdot}$ Depleted region) as depicted using red asterisks in **Fig. 1B**. For statistical analysis, at least 36 indents were made for each condition near the cathode and anode, respectively.

After nanoindentation, the samples were chemically etched in 15 mL of 40% $HNO_3$ with 16 drops of 65% HF for 15 s to estimate the low pre-existing dislocation density. This chemical etching method also reveals the newly generated dislocations by nanoindentation, compared to the pre-existing dislocations (corresponding to the randomly distributed etch pits).

### 2.4 Molecular Dynamics Simulation

MD simulations of the nanoindentation test on single-crystal $SrTiO_3$ with and without pre-inserted oxygen vacancies were performed in displacement-controlled mode. The initial system is constructed with Atomsk [33]. A cubic unit cell is duplicated 100 x 80 x 100 times to form a supercell of 4 million atoms. Periodic boundary conditions are applied along X and Z, and a large vacuum region 200 Å thick is added along the *Y* direction to mimic a free (010) surface. One sample is free of oxygen vacancies, while in the other sample, 10 oxygen vacancies are randomly introduced in the layer below the top surface. Simulations are performed with LAMMPS [34] using the interatomic potential by Pedone et al. [35]. Coulomb interactions are computed with the particle-particle-particle-mesh (pppm) algorithm in the system containing oxygen vacancies; a uniform background charge is added to ensure that the simulation is charge-neutral. Performing MD simulations at finite temperatures introduces noise and uncertainty into the results. To remove thermal effects and isolate the impact of vacancies on nucleation events, we perform molecular statics (so-called "0 K simulations"), where forces on atoms are minimized using the conjugate-gradients algorithm. A spherical indenter of radius *R*=100 Å is placed above the free surface. Iteratively, the indenter is brought down by 0.2 Å, and then all atoms are allowed to relax. Snapshots of the simulation are then analyzed with OVITO [36]. The dislocation extraction algorithm (DXA) is applied to Sr and O atoms to identify dislocation lines, while the centro-symmetry criterion is applied to Sr and Ti atoms to identify atoms belonging to a defective region (free surface, dislocation, stacking fault), and to hide atoms in the perfect environment. This allows for the simultaneous visualization of both partial dislocation lines and the stacking faults between them.

## 3. Results and analyses

### 3.1. Sample color change and defect chemistry tuning

We performed a stoichiometry polarization-induced electromigration experiment as described in **Sec. 2.2** to generate a concentration profile of the vacancy concentration on the undoped single-crystal $SrTiO_3$**.** In **Fig. 2A,** the schematic shows the oxygen vacancy profile after electromigration with high concentrations on the cathode and lower concentrations on the anode as discussed in literature [27, 37] (typical for undoped and acceptor-doped $SrTiO_3$). The oxygen vacancy concentration is about 3 orders of magnitude higher near the cathode compared to the anode [27, 37]. A saturation current of ~4.5 mA was attained at the end of the electromigration experiment.

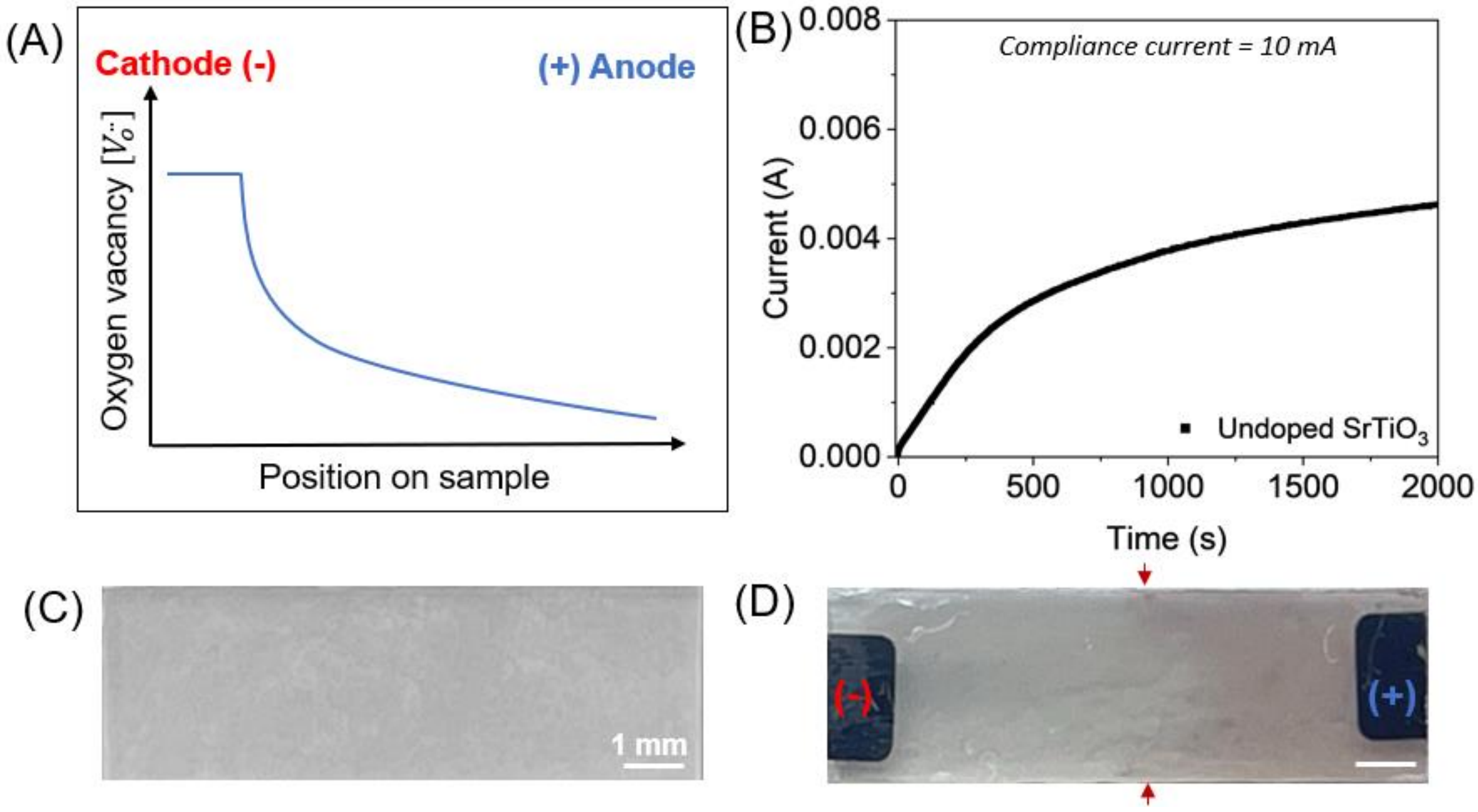

***Fig. 2:*** *(A)Illustration of the oxygen vacancy concentration profile $[V_O^{\cdot\cdot}]$ between the cathode and the anode. A decay profile adapted from [37] at $pO_2$ = 1 bar. (B) The current vs. time plot during electromigration with a current saturation of ~ 4.5 mA with a compliance current of 10 mA. Optical images (C) before, (D) after electromigration. The red arrowheads in (D) indicate the point of confluence of the color fronts between the cathode and anode.*

In **Figs. 2C-D**, the changes in optical appearance (electrocoloration) [28] after electromigration was observed, the confluence point was indicated with red arrows. The initially transparent sample exhibits a reddish-brown color in the anode region and a light grey color in the cathode region. Typically, for undoped $SrTiO_3$, oxygen ions migrate towards the anode (as depicted in **Fig. 1A**). In turn, oxygen vacancies accumulate at the cathode and deplete at the anode. The anode is termed the "oxygen vacancy-depleted region ($V_O^{\cdot\cdot}$ Depleted)" and the cathode "oxygen vacancy-enriched region ($V_O^{\cdot\cdot}$ Enriched)". The change in optical appearance during electromigration is attributed to the change in the electronic state of trace impurity acceptor elements, such as Fe element ($Fe^{4+=Fe^{3++h^{\cdot}}}$) [27, 37]. Note that trace acceptor (Fe) impurities [38] are readily present in commercially undoped $SrTiO_3$.

### 3.2. Incipient plasticity altered by electromigration

To observe the impact of oxygen vacancy concentration gradient after electromigration on the dislocation plasticity, nanoindentation was performed near the two electrodes where the maximum difference in the $V_O^{\cdot\cdot}$ vacancy gradient is expected. Representative load-displacement *(P-h)* curves after electromigration using a spherical tip are presented in **Figs. 3A** (for indents performed as indicated in **Fig. 1B**). The *P-h* responses before the pop-in event are fully elastic irrespective of the position of the indent on the electrodes ($V_O^{\cdot\cdot}$ Enriched or $V_O^{\cdot\cdot}$ Depleted regions). This was fitted using the Hertzian elastic contact theory [39], $P = \frac{4}{3}E_r\sqrt{R}h^{\frac{3}{2}}$, which overlaps with the elastic part of the *P-h* for all tested samples, hinting at an identical elastic response, regardless of oxygen vacancy concentration facilitated by electromigration. *P* is the load, *h* is the penetration depth, *R* is the effective tip radius, and the reduced modulus $E_r$ is defined by $\frac{1}{E_r} = \frac{(1-v_s^2)}{E_s} + \frac{(1-v_i^2)}{E_i}$. $E$ and $v$ are Young's modulus and Poisson's ratio, respectively. Subscripts $i$ and $s$ represents the indenter and sample, respectively. An $E_r$ value of 225 GPa was calculated with $E_i$ = 1140 GPa, $v_i$ = 0.07, $E_s$ = 264 GPa and $v_s$ = 0.237 for $SrTiO_3$ [40].

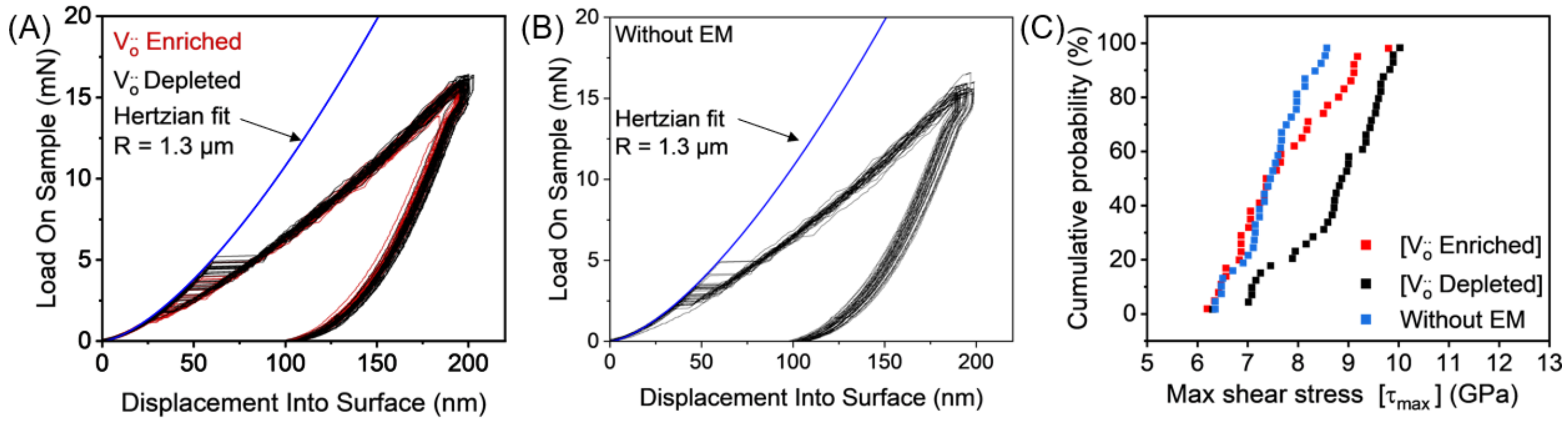

***Fig. 3:*** *(A) Load-displacement (P-h) after electromigration with overlap of the $V_O^{\cdot\cdot}$ Enriched (red) and $V_O^{\cdot\cdot}$ Depleted (black) plots and (B) Load-displacement (P-h) without electromigration (EM). (C) Comparison of the $\tau_{max}$ without electromigration (blue plot) with the $V_O^{\cdot\cdot}$ Enriched and $V_O^{\cdot\cdot}$ Depleted conditions.*

Although the elastic portion of the *P-h* plot overlaps, the load at the pop-in event (corresponding to the onset of dislocation plasticity [41]) correlates with oxygen vacancy concentration as facilitated by electromigration. The average load at the pop-in event is ~ 3.5 mN and ~4.5 mN for the $V_O^{\cdot\cdot}$ Enriched or $V_O^{\cdot\cdot}$ Depleted regions, respectively. For statistical analysis, the cumulative probability distribution of the maximum shear stress ($\tau_{max}$) after electromigration is plotted in **Fig. 3C**. The maximum shear stress at pop-in is estimated using the relation $\tau_{max} = 0.31\left(\frac{6E_r^2}{\pi^3R^2}P_{pop-in}\right)^{\frac{1}{3}}$[39] ( $E_r$, $R$, and $P_{pop-in}$ are the same as stated above). There is a distinct decrease in the maximum shear stress ( $\tau_{max}$ ) in the $V_O^{\cdot\cdot}$ Enriched region compared to the $V_O^{\cdot\cdot}$ Depleted region. The maximum shear stress distribution is confined within a narrow range (~ 6-9 GPa).

To exclude the influence of electromigration and further validate the results in **Fig. 3A**, we repeated the nanoindentation on a different sample (from the same batch to minimize variables) without electromigration experiments. **Figure 3B** shows the *P-h* curve without performing electromigration (without EM). A corresponding cumulative probability of the $\tau_{max}$ (blue dots) is presented in **Fig. 3C** together with the $V_O^{\cdot\cdot}$ Enriched and $V_O^{\cdot\cdot}$ Depleted conditions. The $\tau_{max}$ distribution without electromigration almost coincides with the $\tau_{max}$ in the "$V_O^{\cdot\cdot}$ Enriched" region when compared to the $\tau_{max}$ for the "$V_O^{\cdot\cdot}$ Depleted" region. Detailed discussions of the possible reason for the difference in $\tau_{max}$ is presented in **Sec. 4.1**.

It is important to mention that we can rule out the possible impact of pre-existing dislocations for the observed incipient plasticity of both samples, as depicted in **Fig. 4**. Considering that the estimated average distance between two dislocations is given by $\frac{1}{\sqrt{\rho}}$, where $\rho$ is the dislocation density, for $\rho \approx 10^{10}$ m$^{-2}$ (estimated by counting the dislocation etch-pits per unit area in **Figs. 4A-C),** the average distance between two dislocations is ~10 µm. For both samples, irrespective of the applied electric field, the pre-existing dislocation density remained identical. Considering the effective tip radius of 1.3 µm and the stressed volume underneath the indenter, heterogeneous dislocation nucleation from pre-existing dislocations as possible sources is largely unexpected.

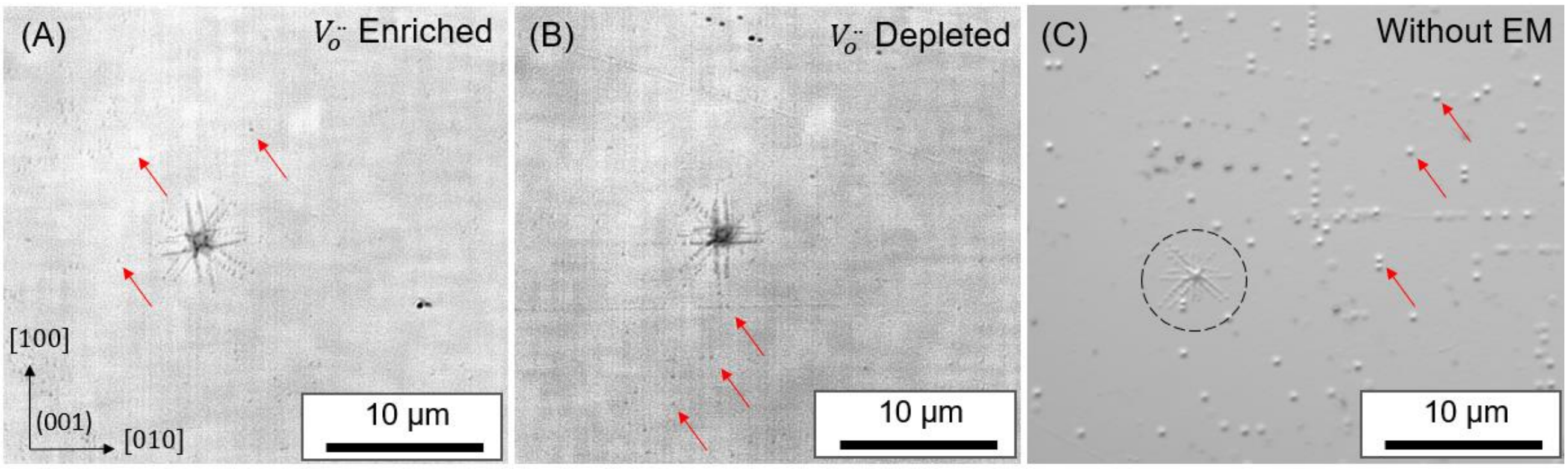

***Fig. 4:*** *Optical images after chemical etching on the electromigration sample (A) and (B), and without electromigration (C). Red arrowheads highlight some of the dislocation etch-pits revealed by chemical etching. The more visible etch-pits in (C) are due to a double chemical etching step (twice the etching time), resulting in larger dislocation etch-pits.*

### 3.4. Molecular dynamic simulation of pristine and oxygen vacancy-rich $SrTiO_3$

To capture the physical picture of the impact of oxygen vacancies on dislocation nucleation, we performed MD simulations on $SrTiO_3$ with and without oxygen vacancies. **Figure 5** shows relevant snapshots from the MD simulations of the nanoindentation process. In the pristine, oxygen vacancy-free $SrTiO_3$ sample, the indenter induces elastic, reversible displacements until it penetrates 12.2 Å below the (010) surface, corresponding to a load of ~2.65 µN (**Fig. 5B**). Then two dislocation half-loops nucleate simultaneously just below the indenter, in two {110} planes normal to one another. Identification with DXA reveals that each loop is made of partial dislocations with Burgers vectors 1/2<110> separated by a stacking fault, as expected from previous studies [42, 43]. As the indenter continues to move, the loops expand below the surface (**Fig. 5C**). In contrast, the oxygen vacancy-rich sample (qualitatively mimicking the $V_O^{\cdot\cdot}$ Enriched region in the experimental result), the initial system contains oxygen vacancies near the free surface (red spots in **Fig. 5D**). In this case, dislocation nucleation occurs when the indenter penetrates only 9.8 Å below the surface, much sooner than in the pristine case. The corresponding load is also smaller, with ~2 μN. Analysis shows that a complete dislocation loop nucleates from an oxygen vacancy (white arrow in **Fig. 5D**) and then expands and dissociates into two 1/2<110> partials (**Fig. 5E-F**).

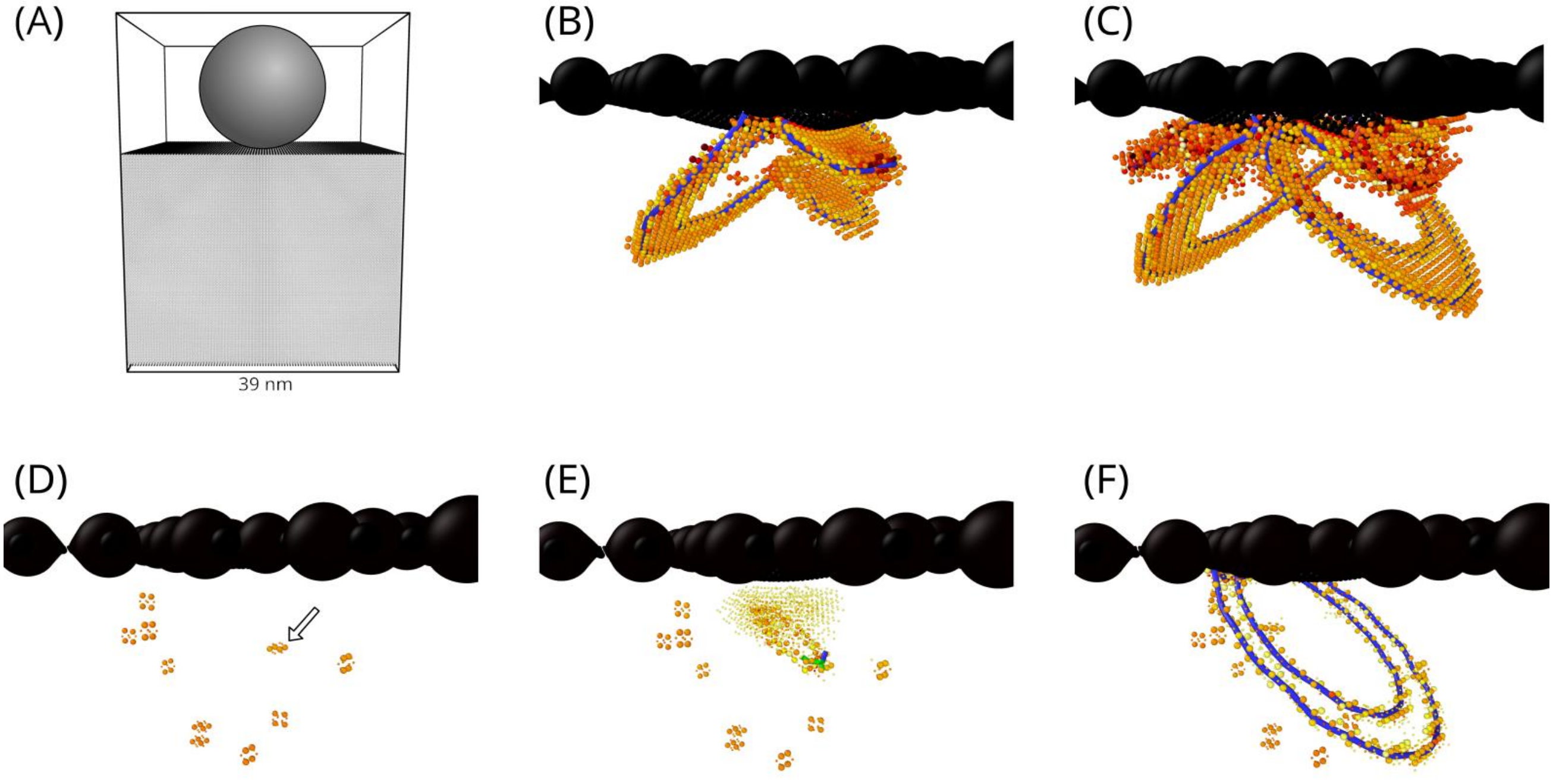

***Fig. 5:*** *Molecular statics simulation of nanoindentation. Atoms in a perfect crystal environment are not displayed to allow visualization of the defects. The top and bottom free surfaces are visible (black atoms), and blue curves indicate the positions of partial 1/2<110> dislocations. (B, C) Snapshots of nanoindentation of pristine (defect-free) $SrTiO_3$ single crystal. Two dislocation loops nucleate below the indenter and propagate (blue curves). (D-F) Snapshots of nanoindentation of $SrTiO_3$ single crystal where oxygen vacancies are initially present below the indenter (red spots in D). A dislocation loop nucleates from the vacancy marked with an arrow in (D), and then propagates.*

The corresponding load-displacement curves for the two simulations are presented in **Fig. 6**. The pop-in event occurred at lower stresses in the simulation cell with $V_O^{\cdot\cdot}$ vacancies, analogous to the experimental result. In the pristine sample, the pop-in event associated with dislocation nucleation occurs when the stress is much higher. This result complements the nanoindentation pop-in results in **Fig. 3**, confirming $V_O^{\cdot\cdot}$ as the defect species promoting dislocation nucleation. We note certain discrepancies between the nucleation stress required to nucleate dislocations during simulation and experiment. These discrepancies could arise from the large difference in tip radius as well as the rate dependence.

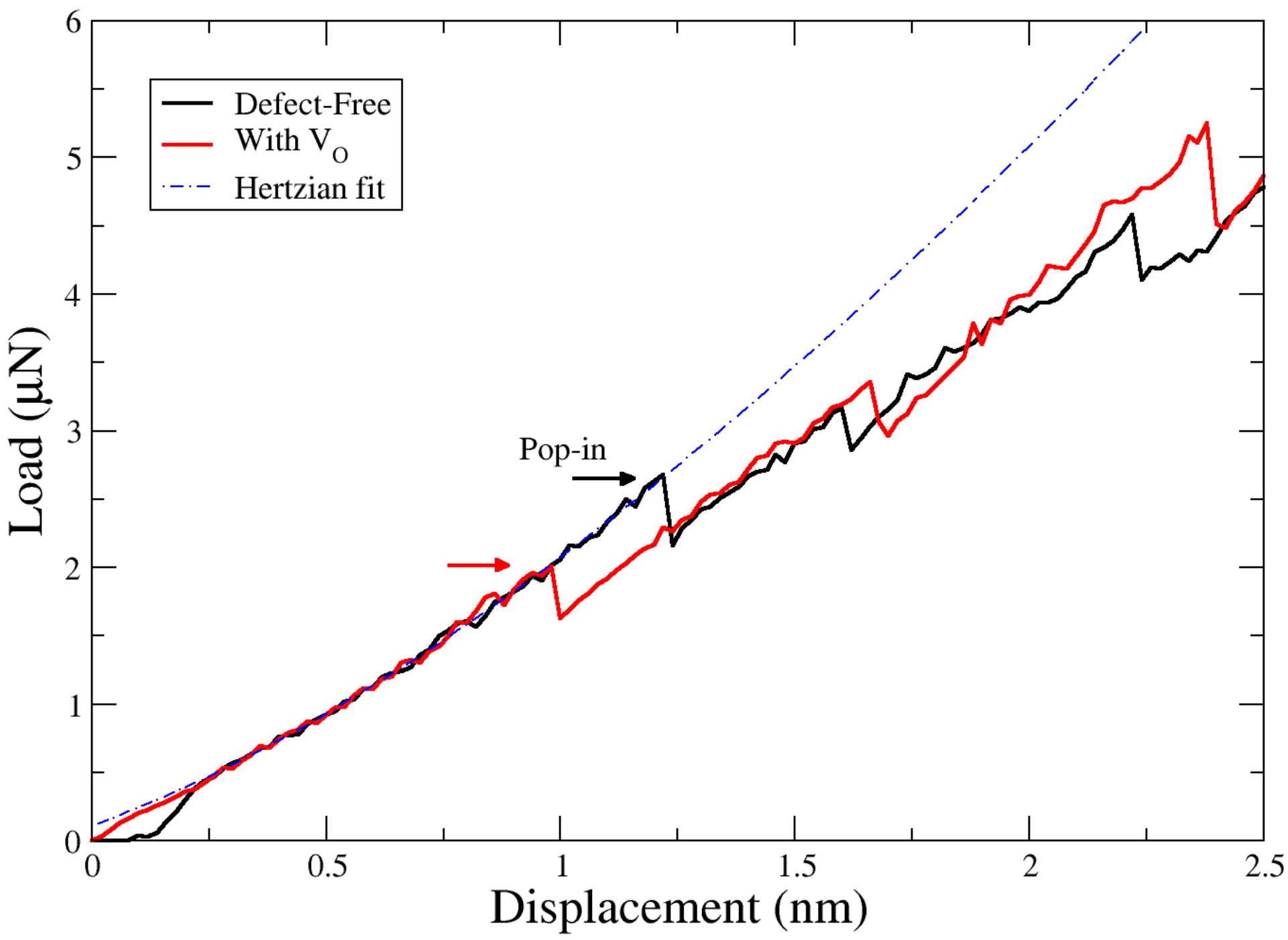

***Fig 6:** Loading curves obtained from molecular static simulations with the pristine sample (black curve) and the one containing oxygen vacancies (red curve). Arrows indicate pop-in events.*

## 4. Discussion

### 4.1. Influence of atmospheric oxygen on the oxygen vacancy concentration

The degree of ionic transfer under an electric field depends on the applied voltage [30], environmental conditions and temperature [27, 30], dopant concentration and doping rate [30, 44], and the transfer rate at the substrate and electrode material interface. **Figure 7** schematically illustrates two possible scenarios likely associated with the electromigration in the $SrTiO_3$ sample. Note that we assume fully ionic-blocking electrode conditions, i.e., only electronic conductivity is possible at the anode and cathode, assured by the electronically conducting platinum electrode. **Figure 7** (left side) shows oxidation and reduction of the sample on the anode and cathode due to the migration of oxygen vacancies. Still, there is no net oxidation/reduction (i.e., the global concentration of oxygen vacancies in the sample remains the same). In this case, the electrodes completely block ionic transfer (only electron-hole conductivity is possible), leading to concentration polarization of the oxygen vacancies [28]. Alternatively, there is an atmospheric contribution leading to re-oxidation of the samples, predominantly at the cathode [27], where re-oxidation is thermodynamically more feasible due to deficient oxygen ion concentration.

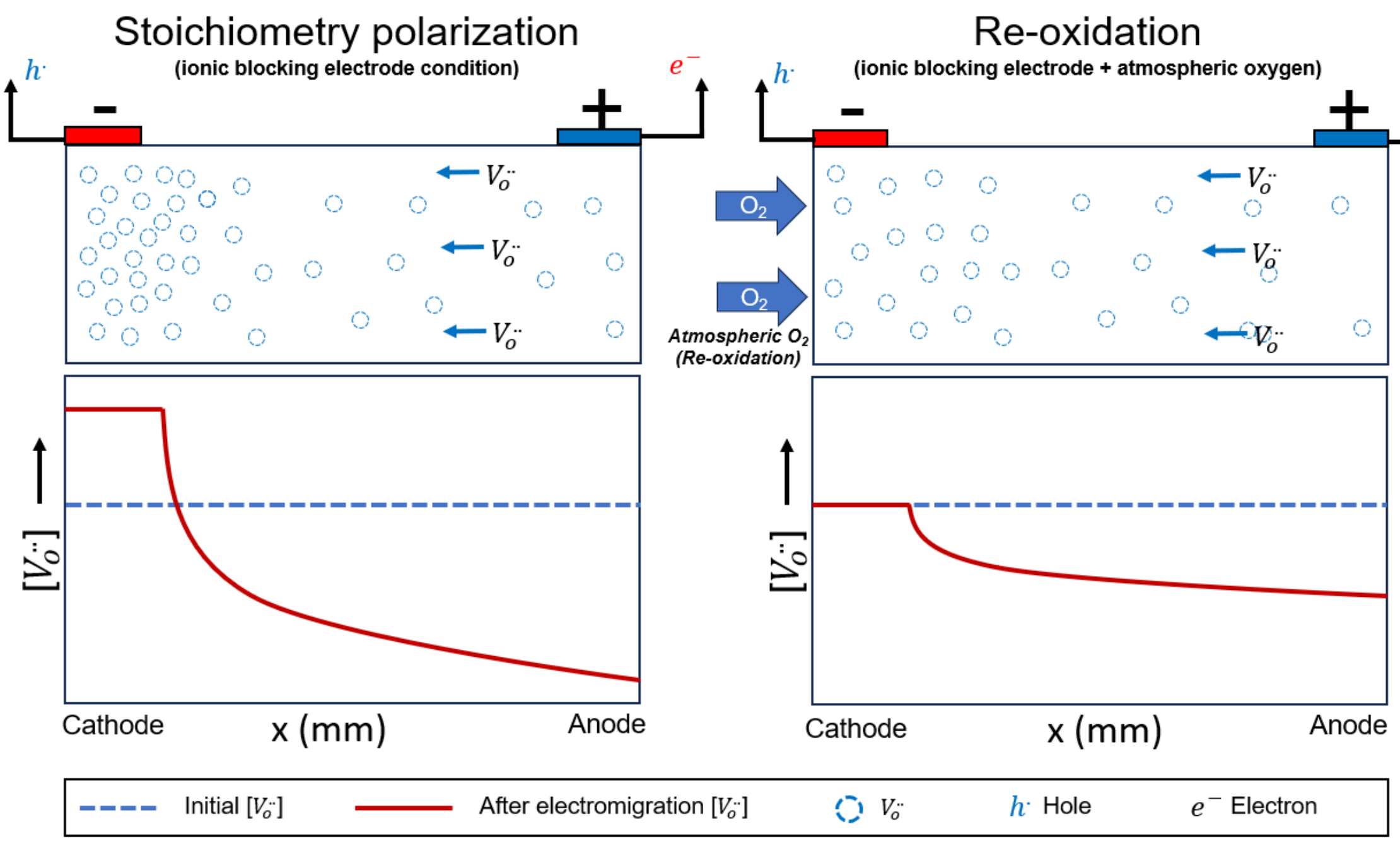

*Fig. 7: Schematic representation of the electromigration setup on a single crystal SrTiO₃ sample depicting a fully ionic (oxygen vacancy exchange) blocking electrode condition (left) and a re-oxidation state superimposed on the stoichiometry polarization-induced electromigration condition (right).*

To identify which condition is associated with our experimental result, we examined the color fronts on the sample after electromigration. The observed dark color front, extending from the anode towards the cathode (**Fig. 2D**), is in part attributed to re-oxidation due to atmospheric oxygen. Alvarez et al. [23] further elucidated this phenomenon by comparing the electromigration test on two different media: air and silicon oil [27]. Tests performed with a sample submerged in silicon oil on Fe-doped (0.05 wt%) $SrTiO_3$ (to prevent atmospheric oxygen ingress) show a shortened dark color front, which is interpreted as indicating a relatively higher oxygen vacancy concentration in the cathode region compared to tests performed in open air. Atmospheric oxygen can readily enter the surface in the cathode region (deficient oxygen atoms), annihilating the $V_O^{\cdot\cdot}$ and restoring the $Fe^{4+}$ state, which leads to an extended dark-colored front for experiments performed in air [27]. Hence, the overall $V_O^{\cdot\cdot}$ concentration is depleted due to the recombination of atmospheric oxygen, as illustrated in **Fig. 7** (right). This re-oxidation at the cathode region is not exclusive to the Fe-doped sample, as the undoped $SrTiO_3$ is readily acceptor-doped due to trace acceptor (Fe) elements [38]. Note that our tests were performed in the open air; therefore, similar to the re-oxidation condition shown in **Fig. 7** (right). Hence, the overall oxygen vacancy concentration in the sample after electromigration is depleted. That means an oxidation state is superimposed on the stoichiometry polarization-induced electromigration, and it is reflected in the $\tau_{max}$ in **Fig. 3**. Ideally, the blue plot (without electromigration) in **Fig. 3C** should lie between the $V_O^{\cdot\cdot}$ Depleted (black plot) and $V_O^{\cdot\cdot}$ Enriched states, but due to surface re-oxidation, there is likely no significant reduction in the $V_O^{\cdot\cdot}$ concentration at the cathode region, resulting in the overlap of the $V_O^{\cdot\cdot}$ Enriched and pristine (*without EM*) states.

### 4.2. Varying tip radius and oxygen vacancy on dislocation nucleation

Oxygen vacancies lead to a local lattice strain, causing a change in the average lattice spacing, which scales with the concentration of $V_O^{\cdot\cdot}$ present in the sample. At high concentrations, $V_O^{\cdot\cdot}$ create *soft* spots in the lattice where lattice bond density is lower, leading to reduced resistance for dislocation nucleation, as reflected in the $\tau_{max}$ in the $V_O^{\cdot\cdot}$ Enriched region in **Fig. 3C**. A similar phenomenon has been reported by the current authors on reduced $SrTiO_3$ [24] and by Jeon et al. [45] on a hydrogenated medium entropy alloy, where hydrogen charging (resulting in the formation of vacancies) was observed to lower the activation energy for the formation of new defects.

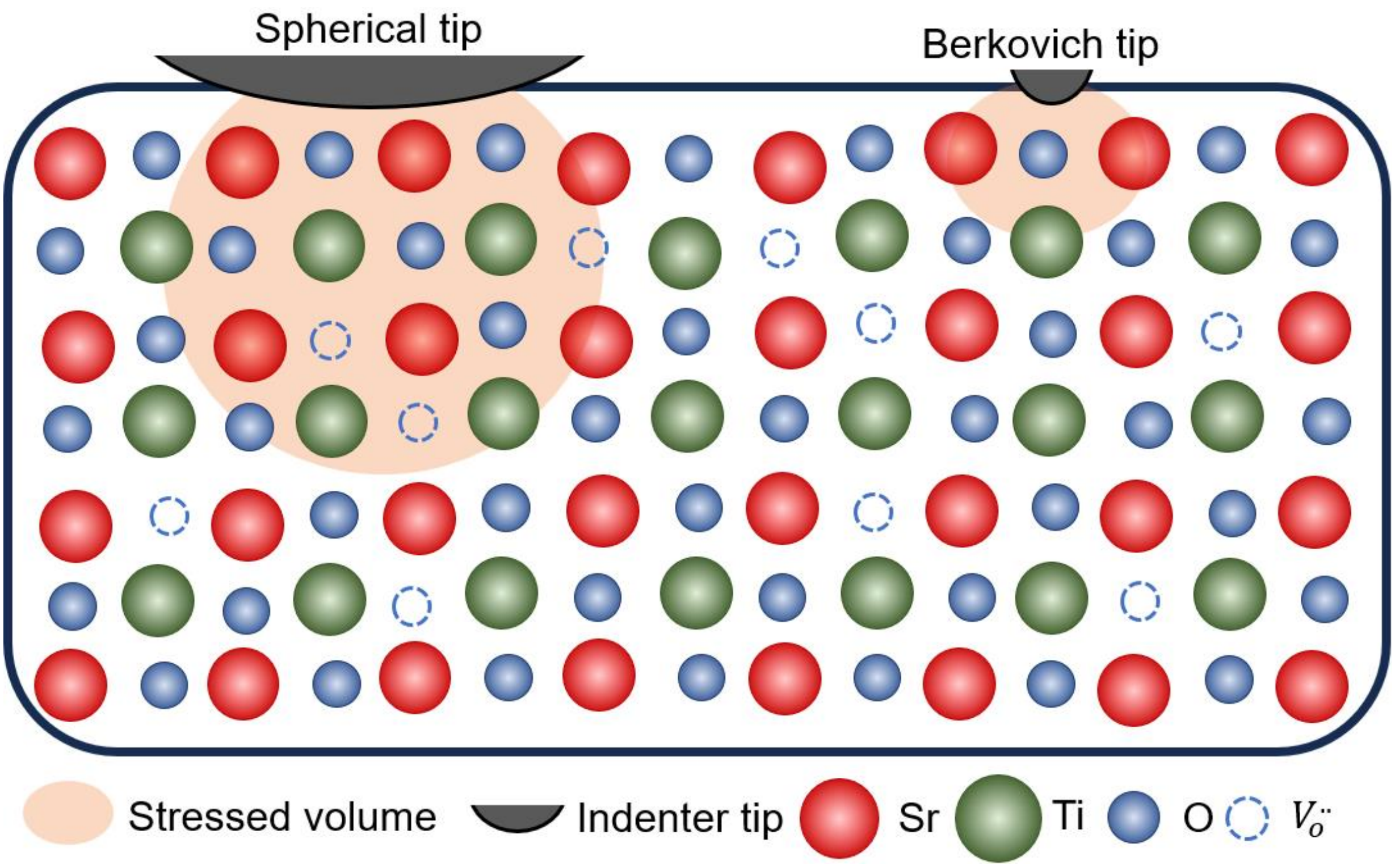

***Fig. 8:*** *2D Schematic illustration depicting the dislocation nucleation with respect to the indenter tip radius. Spherical tip with an effective tip radius of 1.3 µm with oxygen vacancies within the stressed volume (left). Berkovich tip with an effective tip radius of 250 nm and a relatively smaller stressed volume (right). No oxygen vacancy within the stressed volume is reflected in the maximum shear stress for dislocation nucleation after electromigration. Note that the ionic radii are only schematic representations and not in scale with the actual ionic radii, and the indentation contact area is also much smaller than the tip radius.*

An alternative approach to probing the influence of oxygen vacancy concentration on dislocation nucleation can be achieved by varying the indenter tip radius, thereby altering the stressed volume underneath the indenter upon loading. **Figure 8** depicts a schematic illustration of the stressed volume for two different tip radii during nanoindentation. For a given concentration of $V_O^{\cdot\cdot}$, the number of probed $V_O^{\cdot\cdot}$ within the stressed volume depends on the indenter tip size/geometry. Considering only the existence of point defects within the prospective stressed volume before the onset of pop-in. Note there is an estimated low pre-existing dislocation density of ~$10^{10}$ $m^{-2}$ (**Fig. 4**). The population of point defects ($V_O^{\cdot\cdot}$) can be estimated with respect to the stressed volume and point defect ($V_O^{\cdot\cdot}$) density using the equation $n = \rho V$. Where $n$ is the number of $V_O^{\cdot\cdot}$ present in the stressed volume, and $\rho$ is the density of $V_O^{\cdot\cdot}$, and $V$ is the stressed volume, which is defined by[46] $V = \frac{2}{3}\pi r^3 = \frac{2}{3}\pi\left\{R\frac{\pi\tau_{max}}{0.62E_r}\sqrt{\frac{\tau_{max}}{\tau_{critical}}}\right\}^3$, $R$ is 250 nm for the Berkovich tip, and other parameters are as described above. The estimated stressed volume of the sample is ~134 μm$^3$ and ~32.7 μm$^3$ for the spherical and Berkovich nanoindenter tips, respectively. Taking the upper and lower limits of $V_O^{\cdot\cdot}$ as $10^7$ μm$^{-3}$ and $10^4$ μm$^{-3}$ after electromigration [27, 37], there is a one-order-of-magnitude difference in the number of $V_O^{\cdot\cdot}$ present in the probed volume using the spherical and Berkovich nanoindenter tips. Hence, the stress difference for dislocation nucleation from $V_O^{\cdot\cdot}$ is expected to be more evident using a large spherical tip than a sharper Berkovich indenter, as reflected in the $\tau_{max}$ for dislocation nucleation after electromigration in **Fig. 3** (spherical tip) and **Fig. 9** (Berkovich tip), as discussed based on **Fig. 8**.

The maximum shear stress required to nucleate a dislocation using a Berkovich indenter is very close for the $V_O^{\cdot\cdot}$ Enriched and $V_O^{\cdot\cdot}$ Depleted regions. In contrast, ~2 GPa difference in the maximum shear stress is observed using the spherical tip (**Fig. 3C**). We note that both tests were performed on the same sample, and the spherical indents were performed ~24 hours after the Berkovich indents. This is a direct consequence of the indenter size and the probed stressed field near the oxygen vacancies on dislocation nucleation.

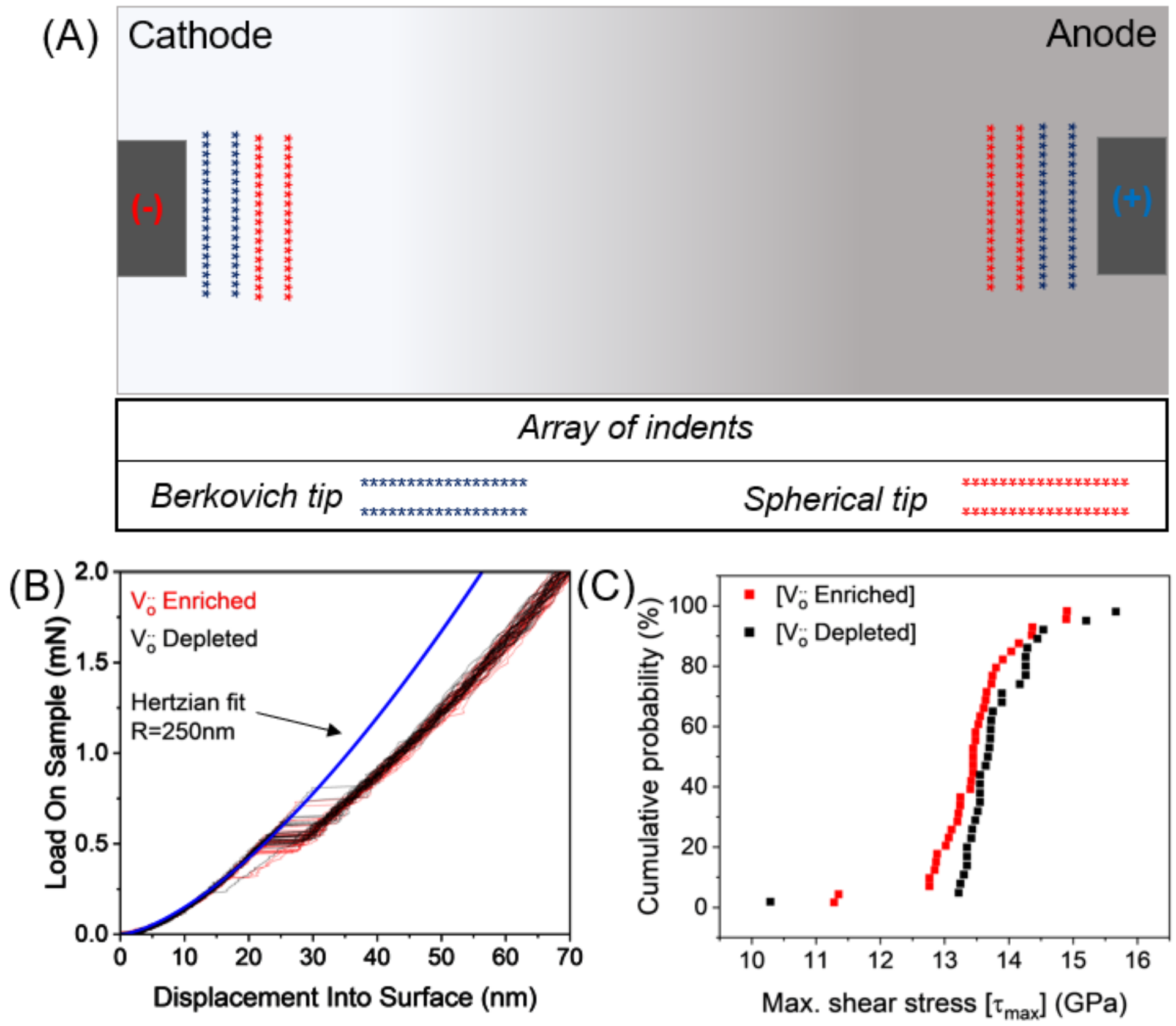

*__Fig. 9:__ (A) Load-displacement plot after electromigration on the sample using a Berkovich tip with an effective tip radius of 250 nm (same sample as in __Fig. 3A__). (B) Cumulative probability of the max. shear stress at the pop-in event for the $V_O^{\cdot\cdot}$ Enriched and $V_O^{\cdot\cdot}$ Depleted regions. (C) Schematic illustration of the position of the indents after electromigration. Blue and red asterisks represent an array of Berkovich and Spherical indents, respectively.*

### 4.3. Beyond electromigration: Doping influence on dislocation nucleation

While this study provides new insights into the impact of oxygen vacancies on dislocation nucleation, several other aspects remain open for further investigation, particularly related to defect chemistry engineering [24]. Doping is one of the most common methods for altering the defect chemistry of oxides, with numerous existing industrial applications. In **Fig. 10**, we highlight the impact of doping on the dislocation nucleation of $SrTiO_3$ during nanoindentation. Analogous to the observation during electromigration, the elastic response before the pop-in event remains unchanged (as shown with the Hertzian fit in **Figs. 10 A-C**), irrespective of the dopants, hinting at an insignificant impact of low-concentration doping (in both cases, 0.05 wt% for acceptor Fe-doping and donor Nb-doping) on the elastic behavior. However, the distinct difference in the pop-in load is highlighted in **Fig. 10D**, showing a higher $\tau_{max}$ required to nucleate dislocations on the Nb-doped (0.05 wt%) sample, while the $\tau_{max}$ overlaps for the undoped and Fe-doped (0.05 wt%) samples.

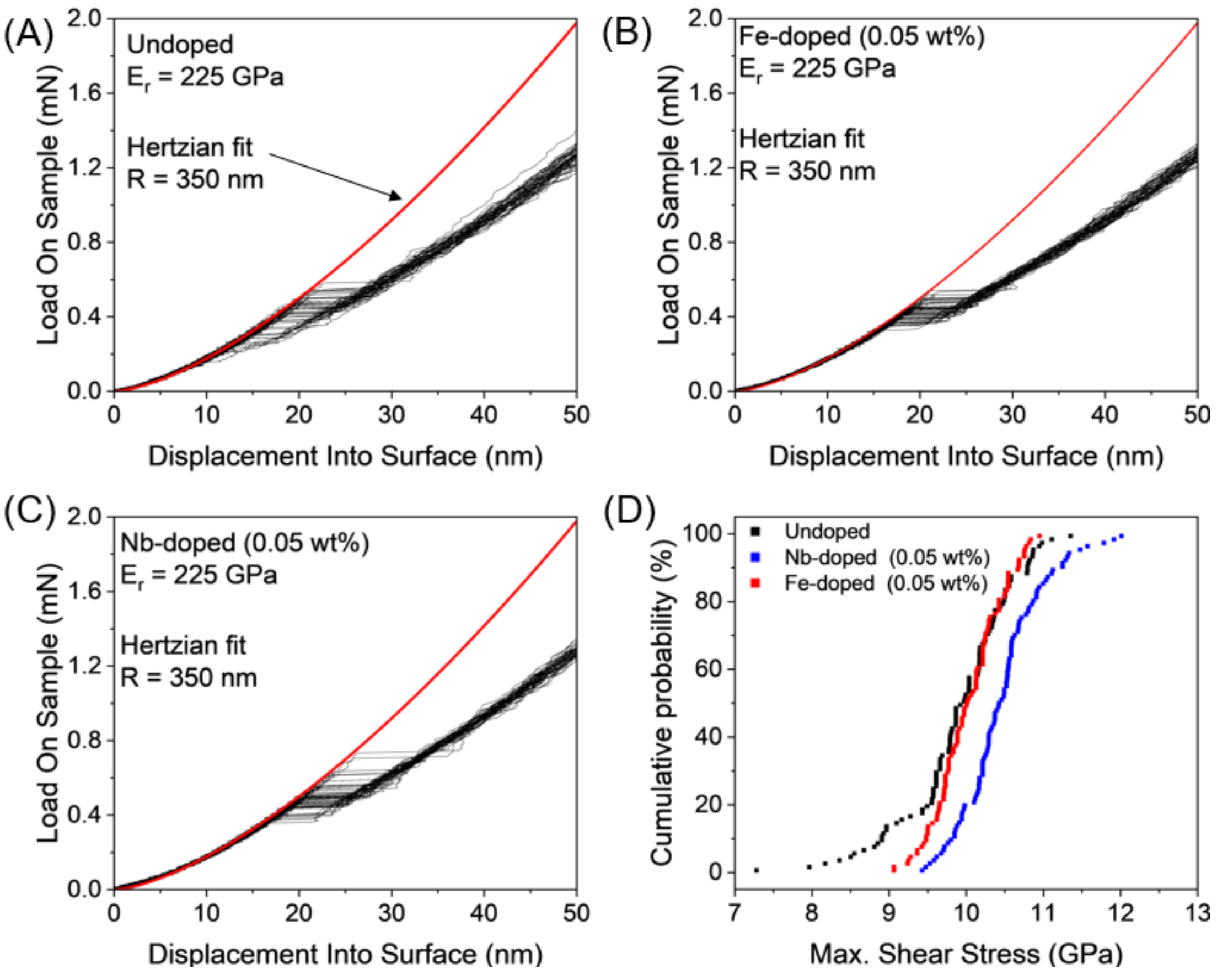

***Figure 10:** Load-displacement plots of (A) undoped, (B) Fe-doped (0.05 wt%), and (C) Nb-doped (0.05 wt%) without electromigration. (D) Cumulative probability of the max. shear stress at the pop-in event for all three samples.*

At ambient temperature and pressure, oxygen vacancies are the dominating point defect species for both undoped and acceptor (Fe)-doped $SrTiO_3$ [32]. In contrast, electrons and/or strontium vacancies dominate for the donor (Nb)-doped $SrTiO_3$, with a much lower oxygen vacancy concentration [17]. These results corroborate our deduction regarding dislocation nucleation promoted by oxygen vacancies, considering the dominating vacancies for the different doped samples. These early insights on the impact of doping on the dislocation plasticity of $SrTiO_3$ (**Fig. 10**) are intriguing and requires further in-depth investigations to consider, for instance, other contributing factors such as the different dopants for possible solid solution hardening, varying doping concentrations and deformation length scales (stressed volume), as reflected in **Sec 4.2**. These aspects are the focus for further studies by the current authors.

## 5. Conclusion

We designed a coupled electromigration-nanoindentation experiment to directly assess the impact of varying point defect concentration on the dislocation nucleation during nanoindentation of single-crystal $SrTiO_3$. Stoichiometry polarization-induced electromigration ensured a spatially distributed gradient in oxygen vacancies on the sample. Nanoindentations with a spherical indenter tip showed a reduction in the pop-in load for high oxygen vacancy concentrations. The favorable dislocation nucleation in the vicinity of oxygen vacancies is likely caused by the local weakening of bonds near oxygen vacancies, as also observed in molecular dynamics (MD) simulations of nanoindentations in both pristine and oxygen vacancy-enriched conditions. The selection of the indenter tip size to probe the oxygen vacancy effect on dislocation nucleation also requires attention, as the size of the stress field and its interaction with defect concentration depend on the tip size and geometry. The larger the stressed volume, the higher the probability of interacting with point defects. Therefore, a more pronounced impact of oxygen vacancy was observed for a tip radius of 1.3 μm compared to a Berkovich tip with an effective tip radius of 250 nm. Similar to the electromigration findings, acceptor doping (Fe-doping) exhibited a comparable trend regarding the impact of oxygen vacancies on dislocation nucleation in $SrTiO_3$. The methodology established in the current study can serve as a valuable tool for understanding the mechanical behavior of functional ceramics under service conditions, particularly when coupled with electric fields, elevated temperatures, and varying oxygen partial pressures.

**Acknowledgment**

C. Okafor, X. Fang, and K. Durst acknowledge the financial support by the Deutsche Forschungsgemeinschaft (DFG, grant No. 510801687). A. Sayyadi-Shahraki expresses gratitude to the Alexander von Humboldt Foundation for its funding. X. Fang is supported by the European Union (ERC Starting Grant, Project MERCERDIS, grant No. 101076167). However, the views and opinions expressed are those of the authors only and do not necessarily reflect those of the European Union or European Research Council. Neither the European Union nor the granting authority can be held responsible.

**Author contributions:**

Conceptualization: XF; Methodology: CO, AS, SB, TF, PH, PC, XF; Investigation: CO, SB, AS, PH; Visualization: CO, PH; Funding acquisition: XF; Project administration: CO, XF; Supervision: XF; Writing – original draft: CO; Writing – review and editing: CO, AS, SB, TF, KD, PH, PC, XF.

**Conflict of interest:**

The authors declare that they have no competing interests.

**Data and materials availability:**

All data are available in the main text.